\documentclass[useAMS,usenatbib]{mn2e}
\usepackage{amsmath,amstext,amsgen,amsbsy,amsopn,amsfonts,theorem}

\usepackage[pdftex]{graphicx}
\usepackage{xspace}
\usepackage{amsmath}
\usepackage{hyperref}
\usepackage{framed}
\usepackage{txfonts}
\usepackage{epstopdf}
\usepackage{gensymb}

\usepackage[normalem]{ulem}

\makeatletter
\def\mr@ignsp#1 {\ifx\:#1\@empty\else #1\expandafter\mr@ignsp\fi}%
\newcommand{\multiref}[1]{\begingroup
\xdef\mr@no@sparg{\expandafter\mr@ignsp#1 \: }%
\def\mr@comma{}%
\@for\mr@refs:=\mr@no@sparg\do{\mr@comma\def\mr@comma{,}\ref{\mr@refs}}%
\endgroup}
\makeatother

\newcommand{\hypref}[2]{\ifx\href\asklfhas #2\else\href{#1}{#2}\fi}
\newcommand{\Secref}[1]{Section~\multiref{#1}}
\newcommand{\secref}[1]{Sec.~\multiref{#1}}

\newcommand{\Tabref}[1]{Table~\multiref{#1}}

\newcommand{\Figref}[1]{Figure~\multiref{#1}}
\newcommand{\figref}[1]{Fig.~\multiref{#1}}
\renewcommand{\eqref}[1]{(\multiref{#1})}

\newcommand{\eq}[1]{\begin{align}#1\end{align}}

\newcommand{\nln}{\nonumber\\}

\title[Tuc III and the LMC]{Modelling the Tucana III stream - a close passage with the LMC}

\author[Erkal et al.]{
\parbox{\textwidth}{
\Large
D.~Erkal$^{1,2}$\thanks{d.erkal@surrey.ac.uk},
T.~S.~Li$^{3,4}$,
S.~E.~Koposov$^{5,2}$,
V.~Belokurov$^{2,6}$,
E.~Balbinot$^{1}$,
K.~Bechtol$^{7}$,
B.~Buncher$^{8,3}$,
A.~Drlica-Wagner$^{3}$,
K.~Kuehn$^{9}$,
J.~L.~Marshall$^{10}$,
C.~E.~Mart{\'\i}nez-V{\'a}zquez$^{11}$,
A.~B.~Pace$^{10}$,
N.~Shipp$^{4}$,
J.~D.~Simon$^{12}$,
K.~M.~Stringer$^{10}$,
A.~K.~Vivas$^{11}$,
R.~H.~Wechsler$^{13,14,15}$,
B.~Yanny$^{3}$,
F.~B.~Abdalla$^{16,17}$,
S.~Allam$^{3}$,
J.~Annis$^{3}$,
S.~Avila$^{18}$,
E.~Bertin$^{19,20}$,
D.~Brooks$^{16}$,
E.~Buckley-Geer$^{3}$,
D.~L.~Burke$^{14,15}$,
A.~Carnero~Rosell$^{21,22}$,
M.~Carrasco~Kind$^{23,24}$,
J.~Carretero$^{25}$,
C.~B.~D'Andrea$^{26}$,
L.~N.~da Costa$^{21,22}$,
C.~Davis$^{14}$,
J.~De~Vicente$^{27}$,
P.~Doel$^{16}$,
T.~F.~Eifler$^{28,29}$,
A.~E.~Evrard$^{30,31}$,
B.~Flaugher$^{3}$,
J.~Frieman$^{3,4}$,
J.~Garc\'ia-Bellido$^{32}$,
E.~Gaztanaga$^{33,34}$,
D.~W.~Gerdes$^{30,31}$,
D.~Gruen$^{14,15}$,
R.~A.~Gruendl$^{23,24}$,
J.~Gschwend$^{21,22}$,
G.~Gutierrez$^{3}$,
W.~G.~Hartley$^{16,35}$,
D.~L.~Hollowood$^{36}$,
K.~Honscheid$^{37,38}$,
D.~J.~James$^{39}$,
E.~Krause$^{28,29}$,
M.~A.~G.~Maia$^{21,22}$,
M.~March$^{26}$,
F.~Menanteau$^{23,24}$,
R.~Miquel$^{40,25}$,
R.~L.~C.~Ogando$^{21,22}$,
A.~A.~Plazas$^{29}$,
E.~Sanchez$^{27}$,
B.~Santiago$^{41,21}$,
V.~Scarpine$^{3}$,
R.~Schindler$^{15}$,
I.~Sevilla-Noarbe$^{27}$,
M.~Smith$^{42}$,
R.~C.~Smith$^{11}$,
M.~Soares-Santos$^{43}$,
F.~Sobreira$^{44,21}$,
E.~Suchyta$^{45}$,
M.~E.~C.~Swanson$^{24}$,
G.~Tarle$^{31}$,
D.~L.~Tucker$^{3}$,
A.~R.~Walker$^{11}$
\\ \emph{(Affiliations can be found after the references)} 
} }

\begin{document}

\label{firstpage}

\maketitle

\begin{abstract}
We present results of the first dynamical stream fits to the recently discovered Tucana III stream. These fits assume a fixed Milky Way potential and give proper motion predictions, which can be tested with the upcoming \textit{Gaia} Data Release 2. These fits reveal that Tucana III is on an eccentric orbit around the Milky Way and, more interestingly, that Tucana III passed within 15 kpc of the Large Magellanic Cloud (LMC) approximately 75 Myr ago. Given this close passage, we fit the Tucana III stream in the combined presence of the Milky Way and the LMC. We find that the predicted proper motions depend on the assumed mass of the LMC and that the LMC can induce a substantial proper motion perpendicular to the stream track. A detection of this misalignment will directly probe the extent of the LMC's influence on our Galaxy, and has implications for nearly all methods which attempt to constraint the Milky Way potential. Such a measurement will be possible with the upcoming \textit{Gaia} DR2, allowing for a measurement of the LMC's mass.
\end{abstract}


\section{Introduction}

One of the unshakeable predictions of hierarchical structure formation in Lambda Cold Dark Matter ($\Lambda$CDM) is that dark matter haloes are triaxial \citep[e.g.][]{bardeen_etal_1986,frenk_etal_1988,dubinski_carlberg_1991,warren_etal_1992,jin&suto_2002,allgood_etal_2006}. Accounting for the dissipation of baryons alters the shapes of haloes, making them more spherical but still significantly flattened and usually triaxial \citep[e.g.][]{dubinski_1994,kazantzidis_etal_2004,debattista_etal_2008,duffy_etal_2010}. The detection and characterization of this triaxiality would represent a stunning confirmation of $\Lambda$CDM and has been the focus of many studies over a wide range of halo mass scales \citep[e.g.][]{Oguri_etal_2005,Corless_etal_2009,evans_bridle_2009,law_majewski_2010}. The unmatched 6D phase space information available around the Milky Way makes it an unprecedented environment for testing halo triaxiality at galaxy scales.

Morphological and dynamical fitting of streams has long been heralded as one of the leading tools to map out the Milky Way halo \citep[e.g.][]{lynden_bell_1982,kuhn_1992,grillmair_1998,zhao_1998,johnston_et_al_1999}. Over the years, this has been attempted with a variety of techniques: comparisons with analytical predictions \citep{ibata_et_al_2001,stream_width}, comparisons with N-body simulations \citep{helmi_2004,johnston_et_al_2005,law_majewski_2010}, orbit fitting \citep{koposov_et_al_2010,hendel_orphan}, Lagrange point stripping methods \citep{gibbons_et_al_2014,bowden_et_al_gd1,kuepper_et_al_2015}, and angle action fits \citep{bovy_pal5_gd1}. While most of these works found results consistent with a spherical halo, \cite{law_majewski_2010} found that a substantially flattened halo was needed to explain the Sagittarius stream \citep{ibata_et_al_2001}. However, all of these works have assumed that the Milky Way is a static potential with no perturbations. 

Several lines of reasoning suggest that the LMC has a substantial mass of $\sim 10^{11} M_\odot$. With a stellar mass of $2.7\times10^9 M_\odot$ \citep{lmc_stellar_mass}, abundance matching gives a peak mass of $2\times 10^{11} M_\odot$ \citep{moster_etal_2013,behroozi_etal_2013}. Proper motion measurements suggest that the LMC and SMC are on their first passage about the Milky Way \citep{kallivayalil_etal_2006a,kallivayalil_etal_2006b,besla_etal_2007}, implying that the current mass is close to the peak mass since the LMC only started being disrupted recently. Note that cosmological simulations also suggest that if the LMC is massive, it is likely on its first approach \citep[e.g.][]{patel_et_al_2017}. Requiring that the SMC is bound to the LMC gives LMC mass on the order of $10^{11}M_\odot$ \citep{kallivayalil_etal_2013}. Furthermore, the plethora of dwarf galaxies and star clusters found near the LMC \citep[e.g.][]{koposov_des_dwarfs,bechtol_des_dwarfs,drlica-wagner_2015,kim_etal_2015,horologiumii_disc,luque_etal_2016} suggest that the LMC has a mass of $\sim 10^{11}M_\odot$ \citep{jethwa_lmc_sats}. Finally, a combination of the timing argument of the Milky Way and M31, as well as the nearby Hubble flow, gives an LMC mass of $\sim 2.5\times 10^{11}M_\odot$ \citep{penarrubia_lmc}. Despite these suggestions of a massive LMC, the only direct mass measurements have been performed in the inner part of the LMC using the dynamics of LMC clusters \citep[$2\times 10^{10}M_\odot$ within 8.9 kpc,][]{Schommer_etal_1992} and the rotation curve of the LMC \citep[$1.7 \times 10^{10}M_\odot$ within 8.7 kpc,][]{vandermarel_lmc}. 

Since the Milky Way mass within the distance to the LMC ($\sim$ 50 kpc) is $\sim 4\times 10^{11}M_\odot$ \citep{deason_etal_2012}, i.e. on the order of the LMC mass itself, it is natural to ask what effect the LMC has had on streams around the Milky Way. Indeed, this possibility was considered in \cite{law_majewski_2010} which used the Sagittarius stream to constrain the potential of the Milky Way halo. They found that the effect of a relatively light LMC, $M_{\rm LMC}<6\times 10^{10}M_\odot$, which was unrealistically fixed to its current position, could have a significant effect on the stream. Accounting for the LMC's orbit, \cite{vera-ciro_helmi_2013} found that a $8\times 10^{10} M_\odot$ LMC can substantially alter the Sagittarius stream and can potentially allow for a very different Milky Way halo than that found in \cite{law_majewski_2010}. More recently, \cite{gomez_et_al_2015} found that a $1.8 \times 10^{11} M_\odot$ LMC would induce a substantial reflex motion in the Milky Way which would alter the shape of the Sagittarius stream. It has also been argued that the LMC may be responsible for the warp seen in the Milky Way's HI disk \citep{weinberg_blitz_2006} and stellar disk \citep{laporte_lmc_effect}.

While almost all of the streams studied in the works mentioned above have been in northern Galactic hemisphere, far from the LMC, it is critical to study streams in the south which may have received a significantly larger perturbation from the LMC. Here we consider the Tucana III (Tuc III) stream which was discovered in the second year of the Dark Energy Survey \citep[DES,][]{drlica-wagner_2015}, with a refined measurement made using DES Year 3 data \citep{shipp_DES_streams}. It is situated at a heliocentric distance of $25$ kpc with a length of $\sim 5^\circ$ (corresponding to a physical length of $\sim$8.4 kpc accounting for projection effects), and is currently $\sim 32$ kpc from the LMC. It joins Pal 5 \citep{pal5disc}, NGC 5466 \citep{ngc5466disc,grillmair_ngc5466}, and possibly the Orphan stream \citep{orphan_disc,grillmair_etal_2015} as the only long, thin streams with known progenitors around the Milky Way. The presence of the Tuc III progenitor makes the stream an ideal candidate for fitting since if the progenitor is not present, its location and velocity must be marginalized over \cite[e.g.][]{bowden_et_al_gd1}. 

With \textit{Gaia} Data Release 2 (\textit{Gaia} DR2) about to give a wealth of exquisite proper motions, parallaxes, and radial velocities for Milky Way stars\footnote{See \url{https://www.cosmos.esa.int/web/gaia/dr2} for more details}, it is critical to understand whether the LMC is indeed exerting a large perturbation on the Milky Way since this will likely affect most methods of inferring the Milky Way potential. In this paper we argue that the Tuc III stream is one such canary in the coal mine. Crucially, we predict that the Tuc III stream has passed within $\sim 15$ kpc of the LMC and that the proper motion of Tuc III depends sensitively on the mass of the LMC. The main effect of the LMC is to induce proper motions perpendicular to the projection of the Tuc III stream on the sky. Thus, if the LMC has had a large effect on the Tuc III stream, the critical signal to look for in \textit{Gaia} DR2 is how well aligned is the stream track of Tuc III with its proper motions. 

This paper is organized as follows. In \Secref{sec:data} we summarize the existing data on the Tuc III stream. In \Secref{sec:tuc3_mw_fit} we describe the stream fitting method  and present fits of the Tuc III stream in the Milky Way potential. Next, in \Secref{sec:fit_lmc} we present dynamical fits of the Tuc III stream in the combined potential of the Milky Way and the LMC and show how the predicted proper motion of Tuc III depends on the mass of the LMC. In \Secref{sec:discussion} we show that \textit{Gaia} DR2 is expected to have sufficient accuracy to detect the effect of the LMC, discuss the limitations of our analysis, and argue that a similar effect should be present in other streams around the Milky Way. Finally, we conclude in \Secref{sec:conclusions}. 

\section{Data on Tuc III} \label{sec:data}

In this section we present measurements of the Tuc III stream necessary for the stream fit. While most of these measurements are taken from the literature, we present a new measurement of the stream track which will be used when fitting the stream. 

\subsection{Stream coordinates and track}

\cite{shipp_DES_streams} report the end points of the Tuc III stream and give the pole corresponding to a great circle fit through these end points. We use this pole to define our stream plane $(\Lambda,B)$, and perform a final rotation so that the progenitor is at the origin. The transformation between $(\alpha,\delta)$ and $(\Lambda,B)$ is given in the appendix of Li et al. in prep. In order to fit the stream we must measure the location of the centroid along the stream (i.e. the stream track). In order to do this, we adopt the following model. We consider six 0.5 degree wide bins  in $\Lambda$ spanning the range of $\Lambda$ from -1.65$^\circ$ to -0.15$^\circ$ and 0.15$^\circ$ to 1.65$^\circ$ (thus avoiding the stream progenitor).  We only consider data very near the stream, $|B|<1^\circ$. In each bin, labeled by $k$, we assume that the density of background stars along $B$ is described by a bilinear
model,  $1+a\Lambda_k+ b B + c\Lambda_k B$, where $\Lambda_k$ is the center of the bin in $\Lambda$ and $a,b,c$ are fitted
constants common for all the bins. The distribution of stream stars across $B$ is described by a Gaussian, $I_k \exp(-0.5(B-B_k)^2/w^2)$, where $B_k,I_k$ are the stream track in $B$ and the stream brightness in a given bin of $\Lambda$, respectively, and $w$ is the global stream width. The resulting posteriori distribution is sampled using a standard ensemble sampler with the following priors, $B_k \sim N(0,0.5), \log(w) \sim U(-\infty, \log(0.2))$ and an improper uniform prior on the stream surface brightness, $\log I_k \sim U(-\infty,\infty)$. As in \cite{shipp_DES_streams}, we select stars using an isochrone from \cite{dotter_etal_2008} with a metallicity of $Z=0.0001$ and an age of 13.5 Gyr. We select all stars within 0.2 in $g-r$ of the isochrone with magnitudes in the range $19 < g < 23$. The resulting stream track is given in \Tabref{tab:tuc3_track} and is well aligned with the stream coordinates. Interestingly, there is no significant offset between the tails on either side of the progenitor, unlike what is seen in the Palomar 5 stream \citep[e.g.][]{pal5disc}.

\begin{table}
\begin{centering}
\begin{tabular}{|c|c|c|}
\hline
$\Lambda$ ($^\circ$)& B ($^\circ$)& $\sigma_B$ ($^\circ$) \\
\hline -1.4 & 0.027 & 0.037 \\
-0.9 & 0.029 & 0.025 \\
-0.4 & 0.002 & 0.109 \\
0.4 & 0.021 & 0.024 \\
0.9 & 0.021 & 0.028 \\
1.4 & 0.053 & 0.060 \\ \hline
\end{tabular}
\caption{Stream centroid (i.e. stream track) in stream coordinates. The stream is well aligned with these coordinates with only small deviations. The final column $\sigma_B$, is the uncertainty on the mean of the fit to the stream and not the stream width. We note that the extent of the stream where we have measured the track is smaller than the measured length in \protect\cite{shipp_DES_streams} since we have limited our analysis to regions where the stream track position is robustly measured.}
\label{tab:tuc3_track}
\end{centering}
\end{table}

\subsection{Distance and distance gradient}

The distance to Tuc III was measured in \cite{drlica-wagner_2015} with a distance modulus of $17.01 \pm 0.16$. We conservatively take the distance to be $25 \pm 2$ kpc. The distance gradient with respect to the angle along the stream, i.e. $\frac{d(m-M)}{d\Lambda}$, was measured by \cite{shipp_DES_streams} with a value of $-0.14 \pm 0.05$ mag/deg. 

\subsection{Radial velocity}

\cite{Simon_TucIII} have measured a radial velocity of $-102.3\pm0.4$ km s$^{-1}$ for the Tuc III progenitor. Li et al. in prep have measured the radial velocities of the Tuc III progenitor and stars along the stream. We choose to only use the results of Li et al. in prep to be more self-consistent since they simultaneously fit the systemic radial velocity and its gradient. In particular, we take the progenitor velocity to be $-101.2\pm 0.5$ km s$^{-1}$ and the radial velocity gradient to be $-8.0 \pm 0.4$ km s$^{-1}$ deg$^{-1}$. 

\section{Fitting Tuc III} \label{sec:tuc3_mw_fit}

Given the data at hand, namely the stream track on the sky, the run of radial velocities along the stream (Li et al. in prep), and the distance gradient along the stream, we can now proceed to fit the stream. Since the Tuc III stream is quite short, we do not currently use it to constrain the potential of the Milky Way. Instead, we fix the Milky Way potential and determine what proper motions, radial velocity, and distance are needed in order to produce the observed stream.

\subsection{Stream generation technique} \label{sec:stream_gen_technique}

In order to perform these fits rapidly, we use the modified Lagrange Cloud stripping (mLCS) technique developed in \cite{gibbons_et_al_2014}. This technique can rapidly generate tidal streams by ejecting stream particles from the Lagrange points of a particle representing the progenitor. In this technique, Tuc III is modelled as a Plummer sphere with a mass of $2\times10^4 M_\odot$ which generates streams with widths similar to Tuc III. We note that while \cite{shipp_DES_streams} estimated a progenitor mass of $8\times 10^4 M_\odot$ using the method of \cite{stream_width}, which relates the stream width to the progenitor mass, that method was derived for streams on near circular orbits and has not been tested on extremely radial orbits. Furthermore, we note that this mass estimation comes with a large uncertainty since the stream width will vary along the stream in a flattened potential \citep[see][for details]{stream_width}. However, since the properties of a stream scale as $m^{1/3}$, where $m$ is the progenitor mass \citep{sanders_binney_2013}, this should not have a large effect on the stream track and radial velocity profile (we have also checked that the results are not very sensitive to the mass). We also assume a scale radius of $10$ pc (although we have checked that the method is largely insensitive to this radius over a realistic range of scale radii). For each set of proper motions, radial velocity, and distance, the progenitor is initialized at the present, rewound for 3 Gyr, and then evolved to the present while disrupting.  

For the Milky Way potential, we choose \texttt{MWPotential2014} from \cite{bovy_galpy}, which satisfies a number of observational constraints. This potential consists of three components: a Miyamoto-Nagai disk \citep{mn_disk}, a power-law bulge with an exponential cutoff, and an NFW halo \citep{nfw_1997}. The Miyamoto-Nagai disk has a mass of $6.8 \times 10^{10} M_\odot$, a scale radius of 3 kpc, and a scale height of 280 pc. We replace the bulge with an equal mass Hernquist bulge \citep{hernquist_1990} with a mass of $5\times 10^9 M_\odot$ and a scale radius of 500 pc for ease in computation. The NFW halo has a virial mass of $8\times10^{11} M_\odot$, a scale radius of 16 kpc, and a concentration of 15.3. For the Sun's motion relative to the local circular velocity we use an offset of $(U_\odot, V_\odot, W_\odot) = (-11.1, 24, 7.3)$ km s$^{-1}$ from \cite{schoenrich_etal_2010} and \cite{bovy_etal_vcirc}. We assume that the Sun is located at a distance of 8.3 kpc from the Galactic center.

In order to make the disruption more physically motivated, we only allow the progenitor to strip stars near pericenter. This is done by recording the pericentric passages when rewinding the orbit and then injecting stream particles at times drawn from a Gaussian (with a constant spread of 0.1 Gyr) about each pericentric passage. Note that we have also tried a spread corresponding to 5\% of of the progenitor's orbital period but there is little difference in the stream. 

In this work we also study how close Tuc III passes to the LMC and how the LMC affects the Tuc III stream. In order to do this, we perform a similar procedure on the LMC as we do for Tuc III. Namely, we rewind the LMC from its present position and then evolve it forward to the present. For the LMC we use proper motions of $\mu_\alpha \cos \delta =  1.91$ mas yr$^{-1}$, $\mu_\delta = 0.229$ mas yr$^{-1}$ from \cite{kallivayalil_etal_2013}, a radial velocity of $-262.2$ km s$^{-1}$ from \cite{vandermarel_lmc_rv}, and a distance of 49.97 kpc from \cite{pietrzynski_lmc_dist}. These values are taken with no uncertainties in our analysis since their errors are much smaller than the uncertainties in the proper motion of and distance to Tuc III. 

\subsection{Priors and the likelihood}

The model contains 4 parameters: the present-day proper motion of Tuc III ($\mu_\alpha \cos \delta, \mu_\delta$), the heliocentric radial velocity of Tuc III ($v_r$), and the distance to Tuc III ($r_{\rm Tuc III}$).  For the proper motions, we take a uniform prior over a wide range ($-10 {\rm \, mas \, yr}^{-1} < \mu_\alpha \cos \delta < 10 {\rm \, mas \, yr}^{-1}, -10 {\rm \, mas \, yr}^{-1} < \mu_\delta < 10 {\rm \, mas \, yr}^{-1})$. For the radial velocity, we use a Gaussian prior of $-101.2\pm0.5$ km s$^{-1}$ based on the results of Li et al. in prep. For the distance, we use a Gaussian prior of $25 \pm 2$ kpc based on the measurement in \cite{drlica-wagner_2015}.

For each generated stream, the likelihood function is defined for the stream track, radial velocity, and distances. The stream is observed from the Sun's location which is assumed to be at $(-8.3,0,0)$ kpc. To measure the stream track in the simulation, we mask around the region within 0.25$^\circ$ of the progenitor (i.e. $
|\Lambda| < 0.25$) and then perform a linear fit for the stream within 1.65$^\circ$ of the progenitor (where the stream is observed). This linear fit is of the form

\eq{ B = m_{\rm track} \Lambda + B_0, }
where $B_0$ is the intercept and $m_{\rm track}$ is the slope of the stream track. This linear fit is then compared against the same fit performed on the observed stream track, accounting for covariance between the slope and intercept of the fit. This gives a log likelihood of

\eq{ \log \mathcal L^{\rm track} = &- \frac{1}{2} \log\big|2 \pi (\mathbf{C} + \mathbf{S} )\big| \nln &- \frac{1}{2} (\vec{\chi}_{\rm obs} -\vec{\chi}_{\rm sim})^{\rm T} (\mathbf{C}+\mathbf{S})^{-1} (\vec{\chi}_{\rm obs} -\vec{\chi}_{\rm sim}), }
where $\mathbf{C}, \mathbf{S}$ are the covariance matrices of the fits to the observed and simulated stream track respectively and $\vec{\chi}_{\rm obs}, \vec{\chi}_{\rm sim}$ are vectors containing the slope and intercept of the fits to the observed stream and simulated stream respectively.

For the radial velocity, a linear fit is performed to the radial velocity within 1.65$^\circ$ of the progenitor, masking out the inner $0.25^\circ$. This fit is of the form,
\eq{v_r = m_{\rm vel} \Lambda + v_0, }
where $v_r$ is the radial velocity, $v_0$ is the intercept, and $m_{\rm vel}$ is the velocity gradient. The gradient is then compared against the radial velocity gradient observed in Li et al. in prep. This gives a log-likelihood of

\eq{ \log \mathcal{L}^{\rm vel} = &-\frac{1}{2} \log{2\pi(\sigma_{m,obs}^2+\sigma_{m,sim}^2)} \nln 
&- \frac{1}{2} \frac{(m^{\rm obs}_{\rm vel}-m^{\rm sim}_{\rm vel})^2}{\sigma_{m,obs}^2+\sigma_{m,sim}^2} ,}
where $\sigma_{m,{\rm obs}}, \sigma_{m,{\rm sim}}$ are the uncertainties of the velocity gradients of the observed and simulated streams respectively, and $m^{\rm obs}_{\rm vel},m^{\rm sim}_{\rm vel}$ are the gradients of the observed and simulated streams respectively. Note that the radial velocity of the progenitor does not appear in the likelihood since it is used in the prior.

These two log-likelihoods are then added to the priors to give the total log-likelihood. Finally, we note that  we also have the distance gradient which is used as an independent check on the fits.

\subsection{Grid search in proper motion} \label{sec:mw_grid}

Before fitting the stream, we evaluated the likelihood on a grid in the proper motions of Tuc III, $(\mu_\alpha \cos\delta, \mu_\delta)$. This search is done to check if there are multiple solutions for the proper motions which can match Tuc III to ensure that that the MCMC results represent the best global fit. The proper motions are varied within the prior range, ($-10 {\rm \, mas \, yr}^{-1} < \mu_\alpha \cos \delta < 10 {\rm \, mas \, yr}^{-1}, -10 {\rm \, mas \, yr}^{-1} < \mu_\delta < 10 {\rm \, mas \, yr}^{-1})$, in steps of $0.1$ mas yr$^{-1}$. For this search, the radial velocity and distance are kept fixed at $-101.3$ km s$^{-1}$ and 25 kpc respectively. We note that many of these proper motions give orbits which are unbound and do not produce any streams since our model assumes that the stream only disrupts near pericenter. This search reveals that for the chosen potential, there is only one region of proper motion which gives a satisfactory fit. Thus this indicates that the fits described in \Secref{sec:mw_fit} are the only solutions for the chosen potential. Note that the results of this grid search are not used in the fits below.

\subsection{Fitting in the Milky Way potential} \label{sec:mw_fit}

\begin{figure*}
\centering
\includegraphics[width=\textwidth]{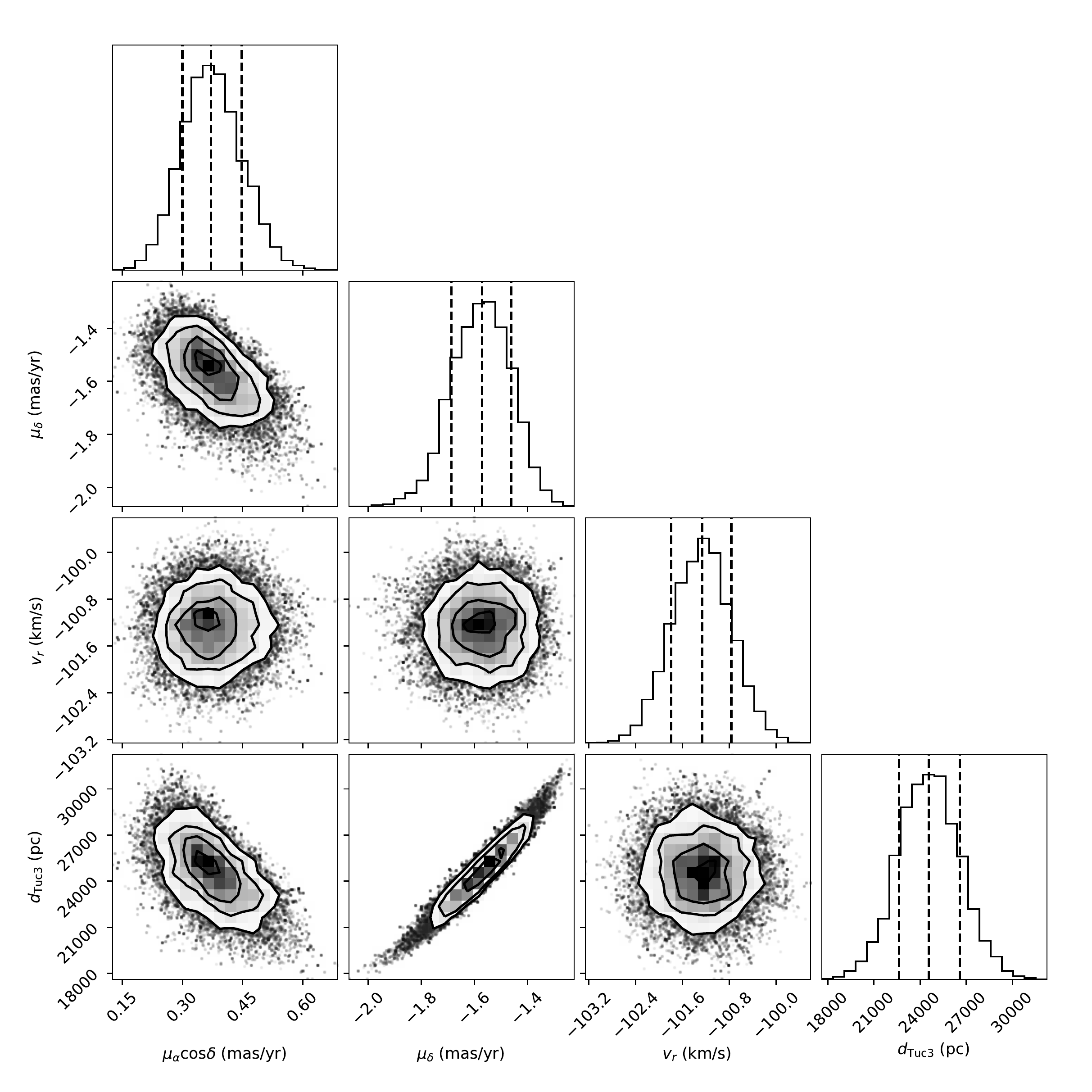}
\caption{Posteriors for proper motions, radial velocity, and distance of Tuc III for fits in the MW. Crucially, the proper motions are tightly constrained by these fits. Note that there is a strong degeneracy between the distance and $\mu_\delta$. This is because the stream has an almost constant declination and thus $\mu_\delta$ must be very close to zero once the reflex motion of the Sun is taken into account. There is also a weaker correlation between $\mu_\alpha \cos \delta$ and $\mu_\delta$ and between $\mu_\alpha \cos \delta$ and the distance to Tuc III. The dashed black lines show the 15.9, 50, and 84.1 percentiles. This figure was made using \texttt{corner} \protect\citep{corner}.} 
\label{fig:posteriors_MW}
\end{figure*}

The fits are done using \texttt{emcee} \citep{emcee}\footnote{\url{http://dfm.io/emcee}}. The walkers are initialized from the priors with the requirement that the progenitor has a pericenter during the simulated disruption so that a stream is produced. We used 200 walkers and 1000 steps (approximately 40 autocorrelation lengths) with a burn-in of 500 steps. \Figref{fig:posteriors_MW} shows the posteriors on the proper motions, radial velocity, and distance. The present-day observations of Tuc III give tight constraints on the proper motion which can be tested with the upcoming \textit{Gaia} DR2 release. Note that there is a strong degeneracy between $\mu_\delta$ and the distance to Tuc III. This is due to the fact that the stream has an almost constant declination and thus $\mu_\delta$ should be very close to zero once the reflex motion of the Sun is taken into account. Improving the distance errors would dramatically improve the uncertainty in $\mu_\delta$. Given this near alignment, the proper motion along the stream is roughly $\mu_\alpha \cos \delta$ minus the Sun's reflex motion. This proper motion has a much larger uncertainty since it still produces a stream aligned with the Tuc III stream for a wide range of values. Thus, improving the distance errors would give little improvement in the uncertainty of $\mu_\alpha \cos \delta$. 

\begin{figure}
\centering
\includegraphics[width=0.45\textwidth]{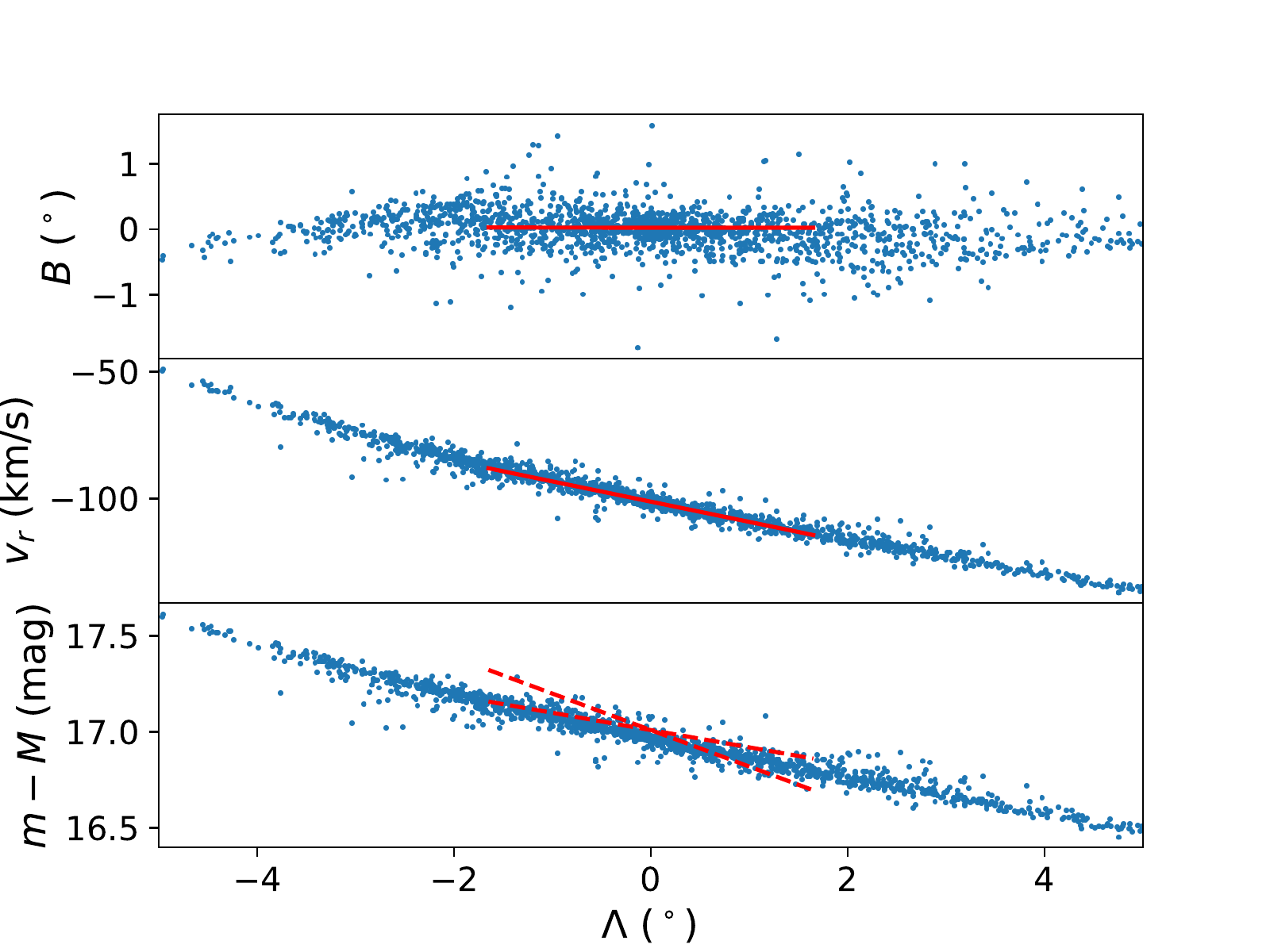}
\caption{Best-fit Tuc III stream in the Milky Way potential. \textbf{Top panel} shows the stream on the sky in stream coordinates. The blue points show the result of the mLCS simulation and the red line shows the best-fit line to the stream track. \textbf{Middle panel} shows the radial velocity along Tuc III. \textbf{Bottom panel} shows the distance modulus along the stream. The dashed red lines shows the 1-$\sigma$ spread of the observed distance modulus gradient, which is provided as an independent check and not used in any of the fits. Note that we only show the stream within $|\Lambda| < 5^\circ$.} 
\label{fig:tuc_MW}
\end{figure}

\Figref{fig:tuc_MW} shows observables for the best-fit Tuc III stream. The model matches the stream track and radial velocity. However, there is a slight difference in the distance gradient which was not used in the fit. Note that the extent of the simulated stream should not be compared against the observed extent since we are using a fixed age for the disruption. Our technique is only meant to reproduce the stream track and radial velocities of the stream, not to match it entirely.

\subsection{Orbit of Tuc III in the Milky Way} \label{sec:orbit_mw}

\begin{figure}
\centering
\includegraphics[width=0.45\textwidth]{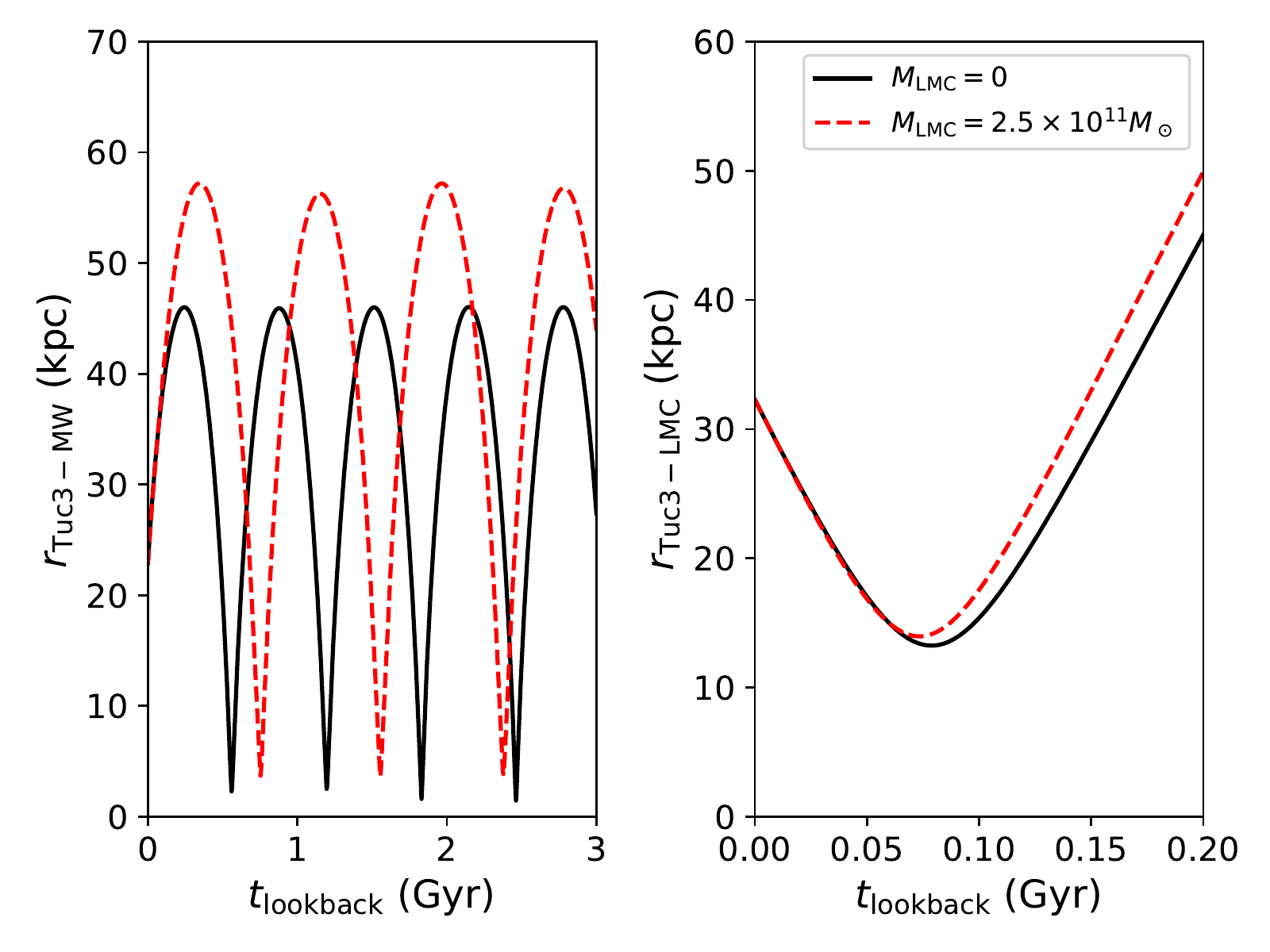}
\caption{Best-fit Tuc III orbit in the Milky Way potential with and without the LMC. The black curve shows the best-fit orbit without the LMC and the dashed red curve shows the best-fit orbit including a $2.5\times10^{11} M_\odot$ LMC. The x-axis on both plots show the lookback time, where $t_{\rm lookback}=0$ is the present. The left panel shows the distance between Tuc III and the MW while the right panel shows the distance between Tuc III and the LMC. Note the close passage between Tuc III and the LMC approximately 75 Myr ago. } 
\label{fig:tuc_orbit}
\end{figure}

\Figref{fig:tuc_orbit} shows the orbit for the best fit to the Tuc III stream. The black curves show the orbit without the LMC and the red curves show the orbit when the LMC is included, which will be discussed in \Secref{sec:fit_lmc}. We find that the orbit of Tuc III around the MW is extremely radial \citep[confirming the claims in][]{Simon_TucIII,shipp_DES_streams}, with an eccentricity of $0.93 \pm 0.01$. The orbit has a pericenter of $1.8 \pm 0.2$ kpc and an apocenter of $45 \pm 4$ kpc (all values are given as the median with 15.9/84.1 percentiles for the errors). The large distance gradient of $0.14\pm0.05$ mag/deg \citep[due to Tuc III's eccentric orbit,][]{shipp_DES_streams} implies that the $\sim$5 degree observed length of the tidal arms corresponds to a physical length of $8.4\pm 2.8$ kpc at a distance of 25 kpc for the Tuc III progenitor, i.e. we are viewing the stream at an angle of 75 degrees from perpendicular.

\begin{figure}
\centering
\includegraphics[width=0.45\textwidth]{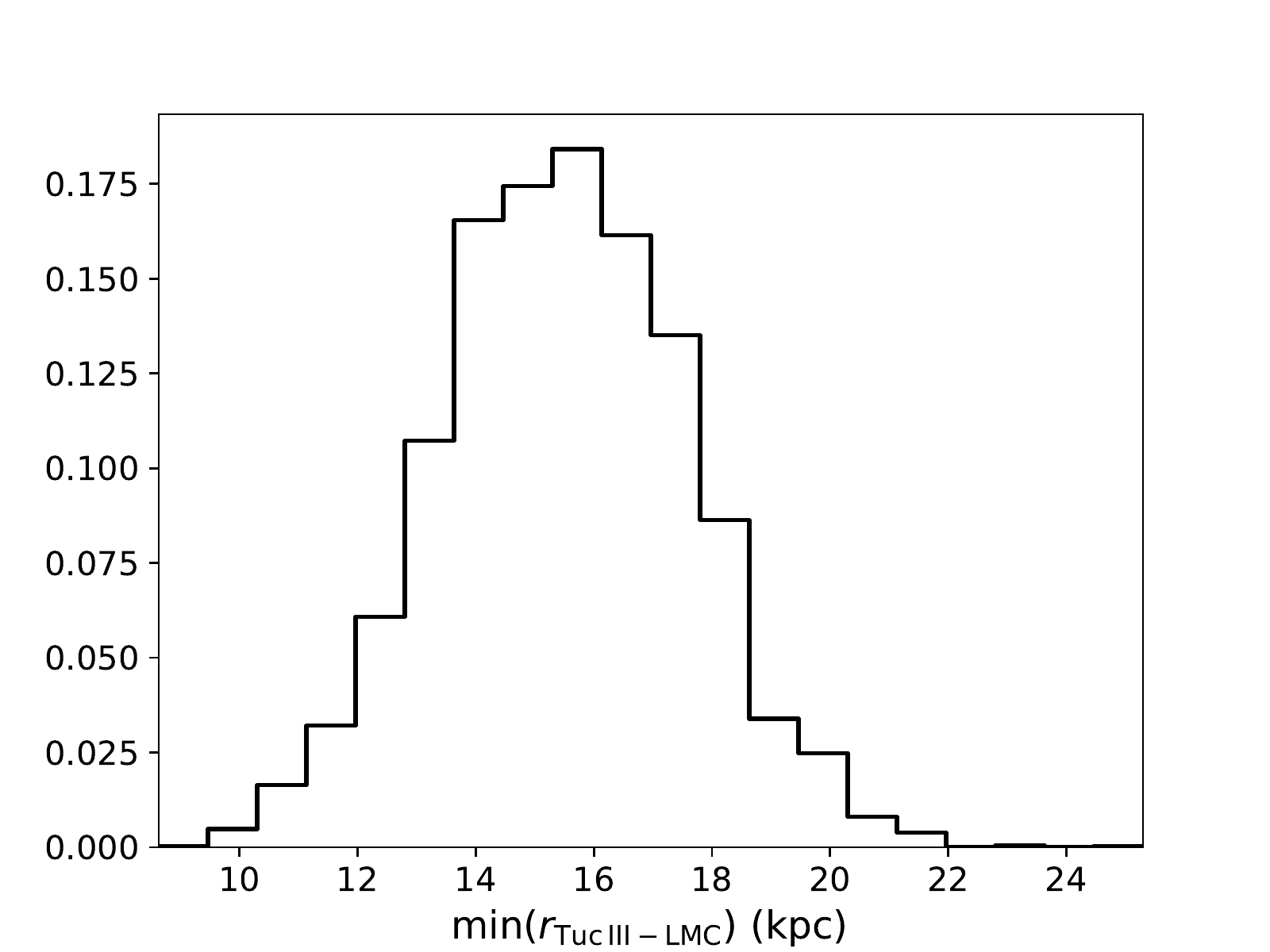}
\caption{Probability distribution of the distance of closest approach between Tuc III and the LMC. This shows that Tuc III likely had a close passage within $15.5 \pm 2$ kpc of the LMC, showing that it is critical to include the effect of the LMC when modelling Tuc III. Note that this figure shows the distance between Tuc III and a massless tracer on the orbit of the LMC so it does not account for the force of the LMC on Tuc III. } 
\label{fig:rmin_dist_MW}
\end{figure}

The right panel of \Figref{fig:tuc_orbit} shows the distance between Tuc III and the LMC. This indicates that there was a close passage between the LMC and Tuc III in the last 100 Myr. For the case with $M_{\rm LMC} = 0$, the LMC is evolved as a massless particle so that Tuc III does not feel any force from the LMC. \Figref{fig:rmin_dist_MW} shows the distribution of minimum distances between Tuc III and the LMC, which indicates that Tuc III likely passed within $15.5 \pm 2$ kpc of the LMC. We note that during this close approach, Tuc III has a large relative velocity, $~300$ km s$^{-1}$, with respect to the LMC. Assuming a relatively light LMC mass of $2\times 10^{10} M_\odot$ \citep[based on observations of its rotation curve at 8.7 kpc][]{vandermarel_lmc}, the force from the LMC during this closest approach is roughly 39\%, 27\% of the force from our Galaxy given the Milky Way potential used in this work \citep[\texttt{MWPotential2014} from][]{bovy_galpy} and \cite{mcmillan_MW_2017} respectively.  This substantial force during closest approach suggests that the LMC likely has a large effect on Tuc III. This is the subject of \Secref{sec:fit_lmc}.

\section{Fitting in the presence of the LMC} \label{sec:fit_lmc}

As mentioned in \Secref{sec:mw_fit}, the best fits for the Tuc III stream pass within 15 kpc of the LMC in the recent past. Given this close passage, it is natural to ask what effect the LMC has on Tuc III. In this section, we include the LMC in the analysis. Note that we do not perform a fit on the LMC mass, but rather we fit Tuc III using different mass LMCs. 

The LMC is modeled as a Hernquist profile \citep{hernquist_1990}. We use 7 different LMC masses of $2,5,10,15,20,25,30\times 10^{10} M_\odot$ with scale radii given in \Tabref{tab:LMC_mass_radius}. For each mass, we pick the scale radius such that the LMC analogue satisfies the rotation curve measurement of the LMC at 8.7 kpc from \cite{vandermarel_lmc}. Dynamical friction is implemented using the approach of \cite{hashimoto_etal_2003}. 

\begin{table}
\begin{centering}
\begin{tabular}{|c|c|}
\hline
$M_{\rm LMC}$ ($10^{10}M_\odot$) & $r_{s, \, {\rm LMC}}$ (kpc)\\
\hline 2 & 0.74 \\ 
5 & 6.22 \\
10 & 12.40 \\ 
15 & 17.14 \\
20 & 21.14 \\
25 & 24.66 \\
30 & 27.85 \\
\hline 
\end{tabular}
\caption{Masses and scale radii of the 7 LMC models used in this work. The LMC is modelled as a Hernquist profile and the scale radii are chosen such that the mass enclosed within 8.7 kpc matches the constraint from \protect\cite{vandermarel_lmc}. }
\label{tab:LMC_mass_radius}
\end{centering}
\end{table}

\subsection{Grid search in proper motions in presence of the LMC}

Before performing the fit, we first perform a grid search in the proper motions as in \Secref{sec:mw_grid}. More concretely, we fix the present-day distance to Tuc III at 25 kpc and the radial velocity at -101.2 km s$^{-1}$, and do a grid search in $\mu_\alpha \cos \delta$ and $\mu_\delta$. This is done twice, with an LMC mass of $1.5\times 10^{11} M_\odot$ and $2.5\times 10^{11} M_\odot$, using the proper motions, distance, and radial velocity for the LMC specified in \Secref{sec:stream_gen_technique}. This grid search reveals only one locus of solutions with likelihoods similar to the best-fit solutions from \Secref{sec:tuc3_mw_fit} suggesting that the results presented below are the only solution in the given Milky Way potential plus an LMC. 

\subsection{Fitting Tuc III with the LMC}

\begin{figure}
\centering
\includegraphics[width=0.45\textwidth]{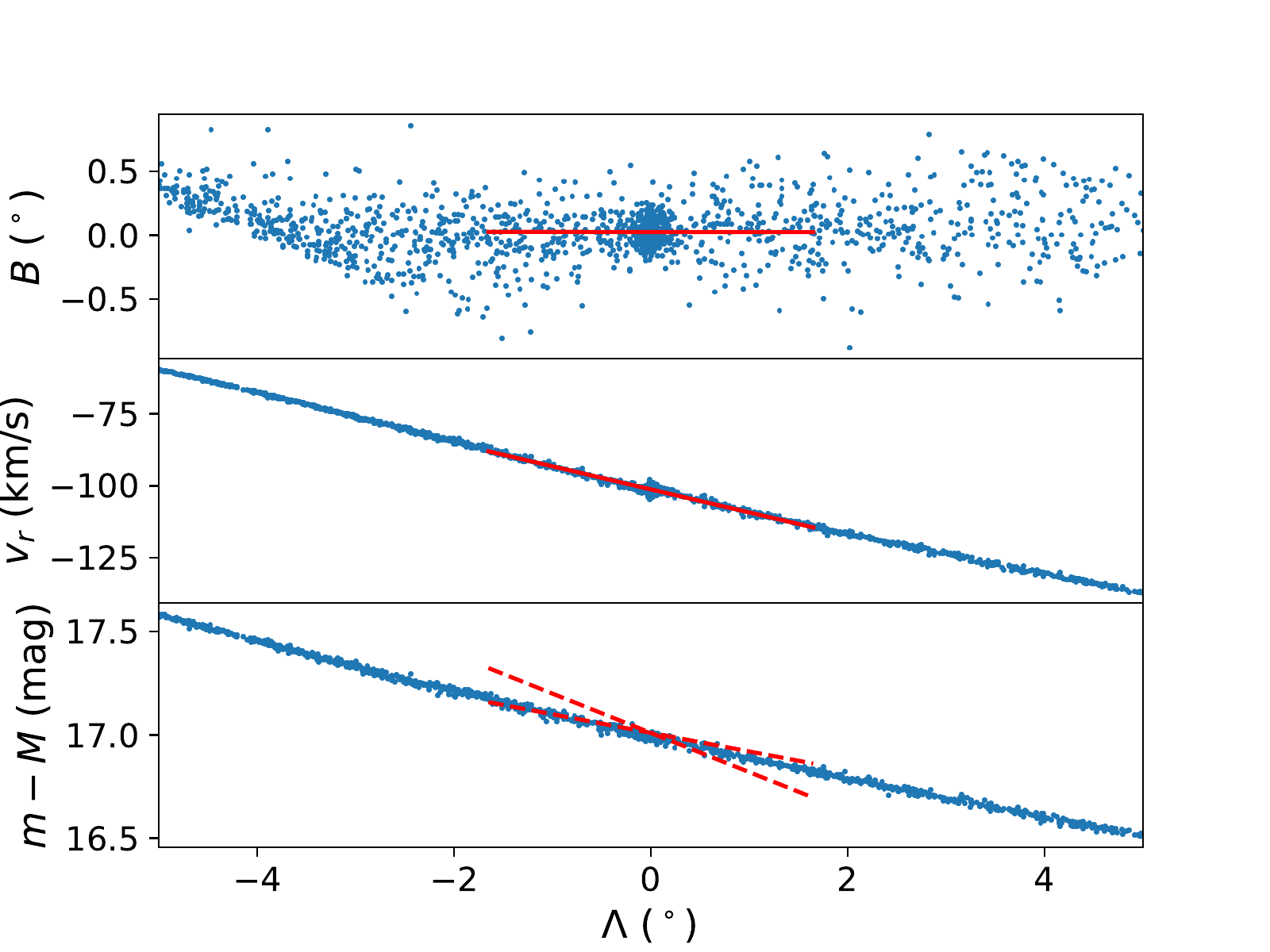}
\caption{Best-fit Tuc III stream in the combined potential of a $2.5\times10^{11}M_\odot$ LMC and the Milky Way potential. The panels are the same as in \protect\figref{fig:tuc_MW}. The stream matches the data very well and is indistinguishable from the best-fit stream in the MW potential. The diagonal truncation towards negative $\Lambda$ is partially due to a projection effect where the stream becomes more radially oriented on the left. It is also likely due to the correlations in energy and angular momentum in the stream debris \citep[e.g.][]{gibbons_et_al_2014}. } 
\label{fig:tuc_2p5e11}
\end{figure}

The fitting works much the same as in \Secref{sec:mw_fit} except that now both Tuc III and the LMC are rewound from their current positions, after which Tuc III disrupts in the presence of the MW and the LMC. For each value of the LMC mass we get a constraint on the proper motions, radial velocity, and distance as in \Figref{fig:posteriors_MW}. We note that there is no qualitative difference in how the posteriors appear so we do not include a corner plot for any of the fits including the LMC. In \Figref{fig:tuc_2p5e11} we show observables for the best fit to Tuc III when a $2.5\times10^{11} M_\odot$ LMC is included. The observables look similar to the best fit with no LMC (see \figref{fig:tuc_MW}) with the exception of a sharp feature in the stream track at $\Lambda < -3^\circ$. However, given the current observations, there is no way to distinguish these two cases. 

\begin{figure*}
\centering
\includegraphics[width=\textwidth]{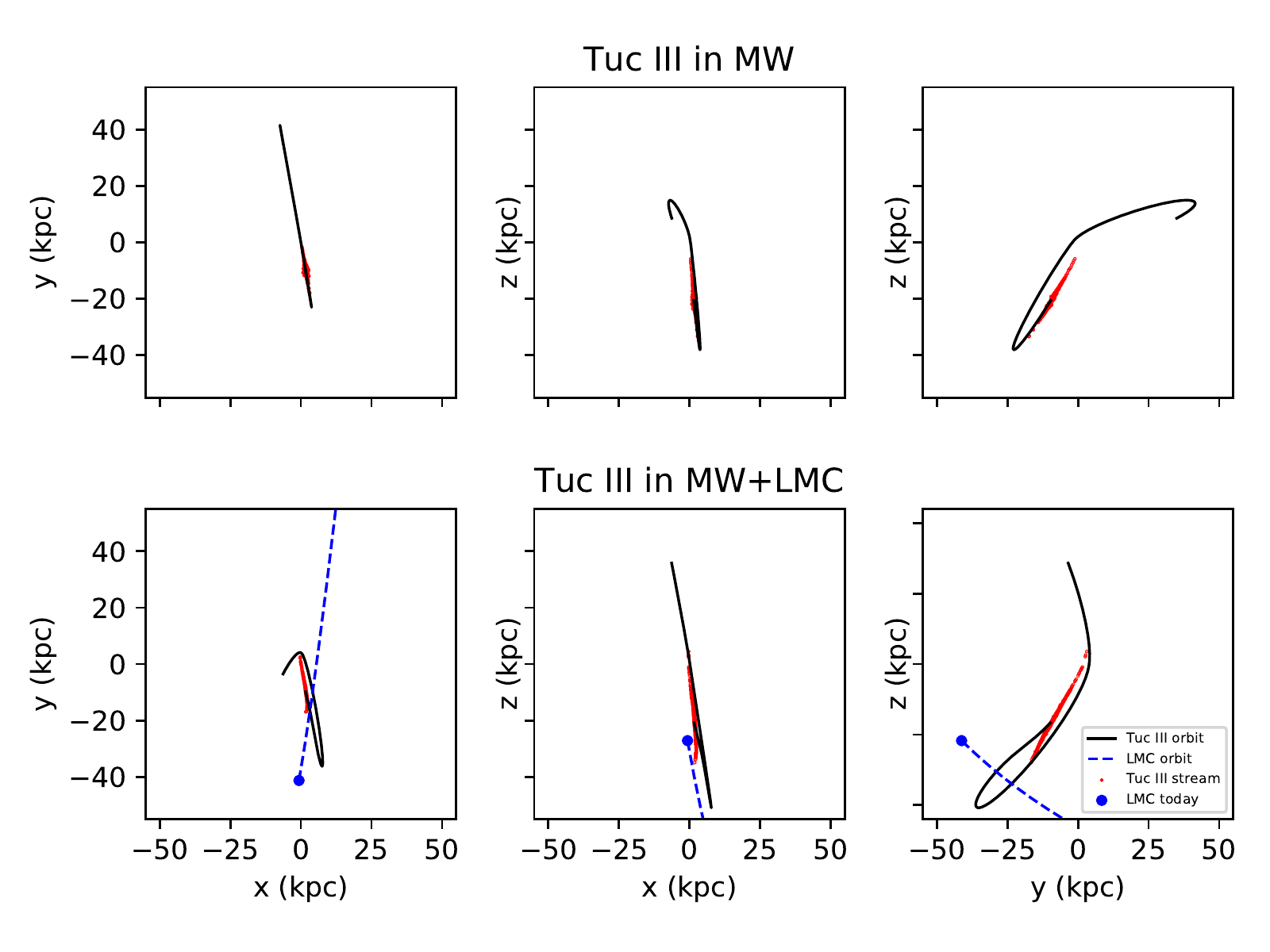}
\caption{Comparison of Tuc III orbit and stream in the Milky Way potential and in the combined influence of the Milky Way and the LMC. \textbf{Top panels} show the best-fit Tuc III stream in the Milky Way potential. Left to right the panels show the  $xy$, $xz$, and $yz$ projections of Tuc III's orbit (black line) and of the Tuc III stream (red points). To make the figure more legible, we only show the orbit of Tuc III over the past 1 Gyr. Note that the stream is well aligned with the final segment of the orbit (e.g. top right panel). \textbf{Bottom panels} show the best-fit Tuc III stream in the combined influence of the Milky Way and a $2.5\times10^{11}M_\odot$ LMC. As with the top panels, the black curve shows the orbit of Tuc III and the red dots show the Tuc III stream at the present. The dashed-blue curve shows the orbit of the LMC and the blue point shows the current location of the LMC. Note that the stream is now significantly misaligned with the final segment of the orbit (e.g. bottom right panel). Since the orbit represents the motion of the stream's progenitor, this means that the progenitor velocity in the two cases should be different. Given that we have fixed the progenitor's radial velocity, this implies that the proper motions should be different in the two cases.} 
\label{fig:comparison_figure}
\end{figure*}

To make the comparison between these streams clearer, we show projections of their orbits and the final streams in \Figref{fig:comparison_figure}. The top panels of this figure show the best-fit Tuc III stream in the presence of the Milky Way while the bottom panels show the best-fit stream in the combined presence of the Milky Way and a $2.5\times10^{11}M_\odot$ LMC. The bottom panels also show the most recent segment of the LMC's orbit to show how close it passes to the stream. Crucially, notice that while the stream is aligned with the final segment of the orbit when evolved in just the Milky Way (e.g. top right panel), there is a significant misalignment when the stream is evolved in the presence of the LMC (bottom right panel). On a similar note, although the streams look similar, the orbits are very different. Since the orbit represents the motion of the stream's progenitor, this implies that the velocities of the progenitor should be different in the two cases. Since both streams have the same radial velocity by construction (due to our priors) this implies that the proper motions must be different.

\begin{figure}
\centering
\includegraphics[width=0.45\textwidth]{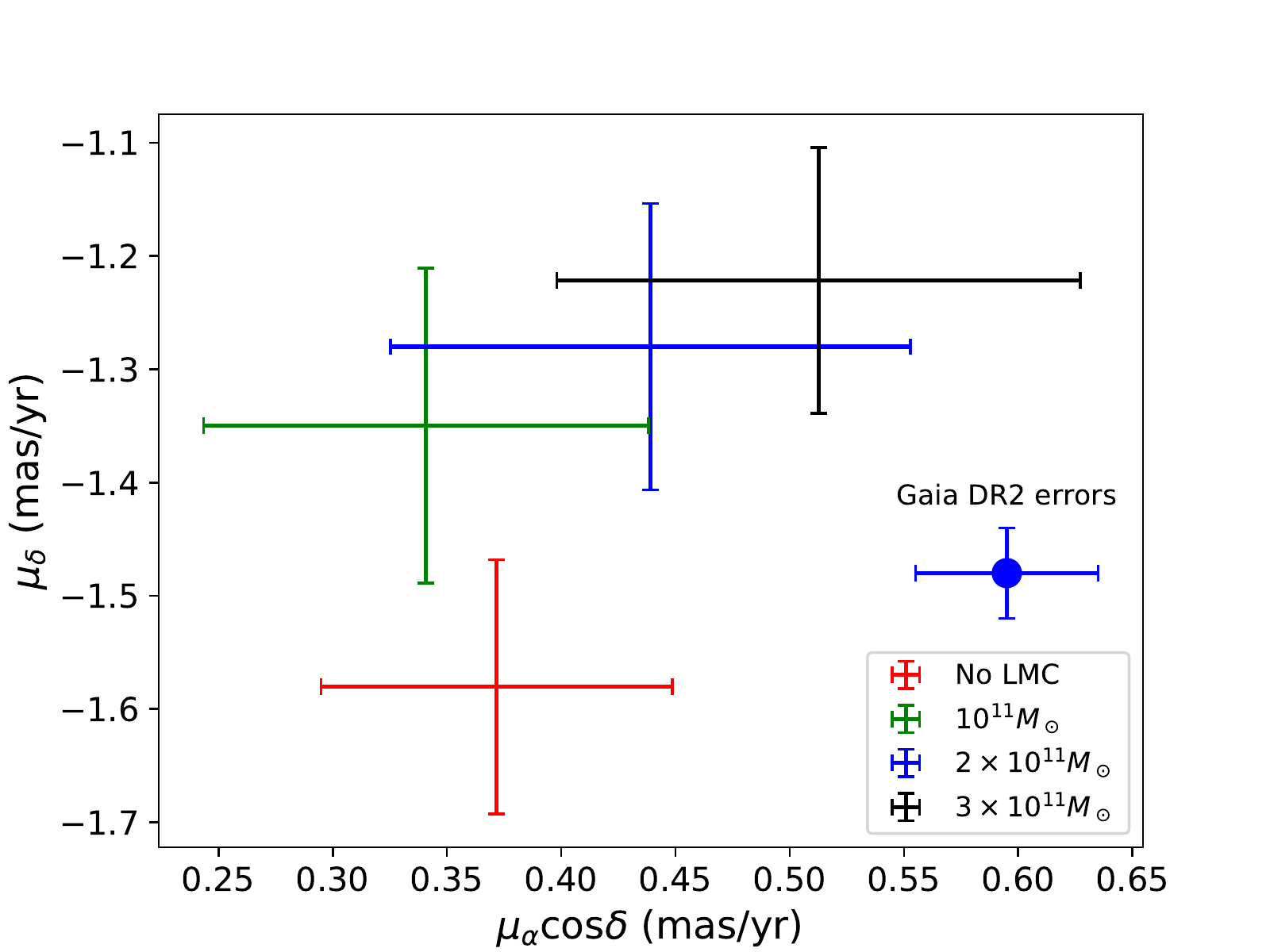}
\caption{Proper motion predictions for Tuc III assuming various masses for the LMC. The large blue point with error bars on the right shows the uncertainty on Tuc III's proper motion of 0.04 mas yr$^{-1}$ expected from \textit{Gaia} DR2 (see \protect\secref{sec:disc_pm}). This shows that the LMC has the largest effect on $\mu_\delta$. Interestingly, since the stream has a declination which is roughly constant, this direction is perpendicular to the stream. Thus the presence of the LMC induces a proper motion which is not aligned with the stream orientation. This is better seen in \protect\figref{fig:tuc_pm_prediction_LB} where the proper motion is given in coordinates aligned with the stream.   } 
\label{fig:tuc_pm_prediction}
\end{figure}

With the stage set, we now explore the proper motion predictions of these models. \Figref{fig:tuc_pm_prediction} shows the proper motions predicted for Tuc III with three different mass LMCs. Interestingly, there is a large change in the proper motions, suggesting that it will be possible to see this effect with \textit{Gaia} DR2 and thus infer the strength of the interaction between Tuc III and the LMC. The main effect of the LMC is to change $\mu_\delta$; there is a relatively minor change to $\mu_\alpha \cos \delta$. Fortunately, the direction of $\mu_\delta$ is roughly perpendicular to the stream track since the stream has an almost constant declination. This means that the LMC induces a significant proper motion perpendicular to the stream track. This is because the close passage with the LMC exerts a large torque on the stream that twists the rotates the stream away from its initial orientation as can be seen in \Figref{fig:comparison_figure}\footnote{See \url{https://youtu.be/qbTgS_TytSc} for a movie showing how dramatically the stream is twisted by the close LMC passage.}. 

\begin{figure}
\centering
\includegraphics[width=0.45\textwidth]{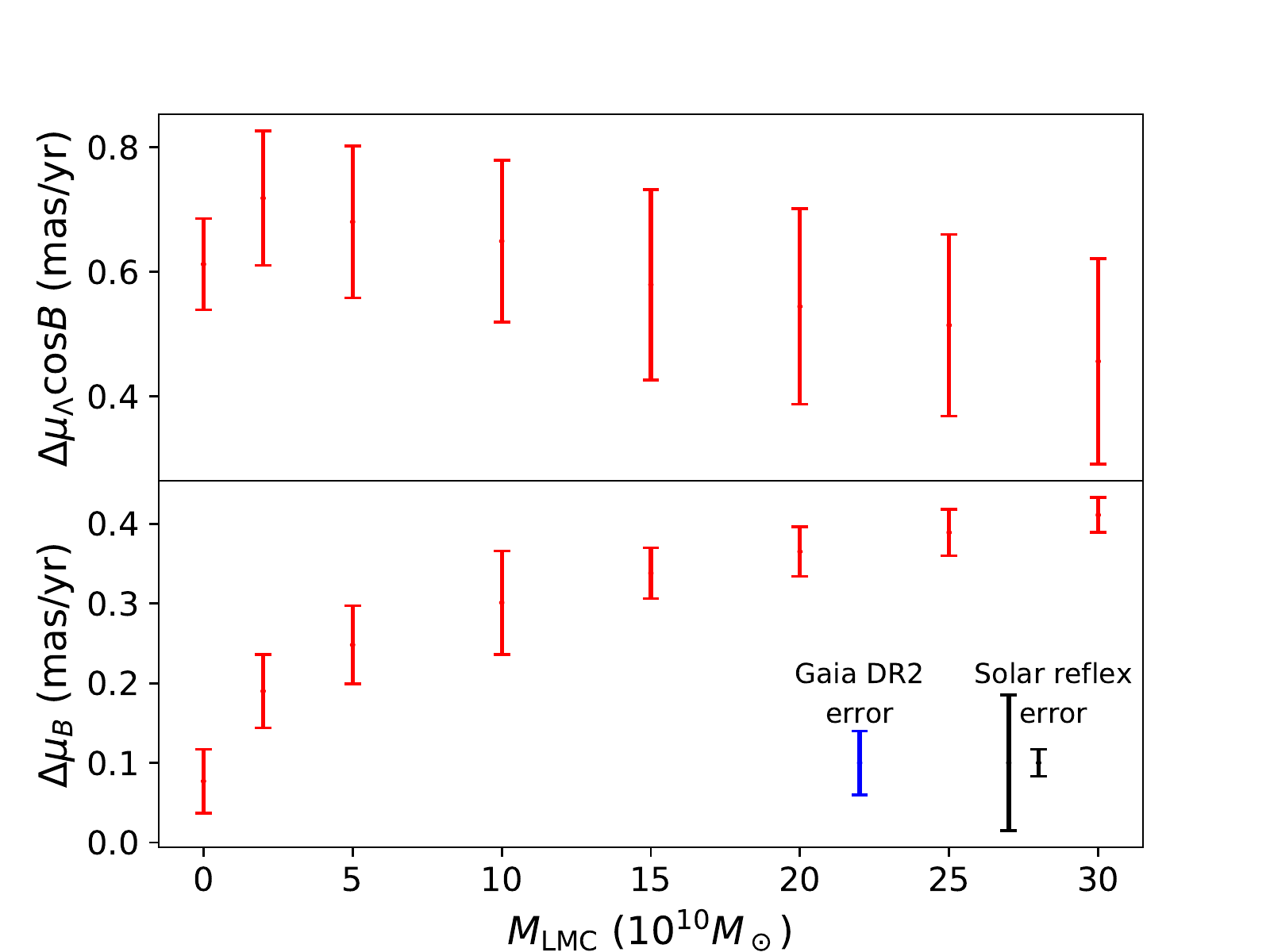}
\caption{Best-fit proper motion prediction for Tuc III in coordinates aligned with the stream versus the LMC mass. Note that we have corrected for the Sun's reflex motion (using the true distance for each realization) from these proper motions and hence denote these as $\Delta \mu_\Lambda \cos B$ and $\Delta \mu_B$. The blue error bar shows the expected uncertainty on Tuc III's proper motion from \textit{Gaia} DR2 while the black error bars show the uncertainty in the solar reflex proper motion assuming a 2\% and 10\% distance uncertainty (see \protect\secref{sec:disc_pm} for more on these error bars). While the proper motion in $\Lambda$ is almost independent of the LMC mass, the proper motion in $B$ depends quite sensitively on it. Thus a measurement of a substantial proper motion in $B$ with {\it Gaia} DR2 can be used to infer the mass of the LMC.   } 
\label{fig:tuc_pm_prediction_LB}
\end{figure}

To further elucidate this point, we show the proper motions in coordinates aligned with the stream in \Figref{fig:tuc_pm_prediction_LB}. In this figure, we have corrected for the Sun's reflex proper motion (using the true distance for each realization) so that the proper motions indicate peculiar motion on the sky. If the LMC has no effect on Tuc III (i.e. $M_{\rm LMC} = 0$), the proper motions are almost aligned with the stream track. However, if the LMC is massive, there will be a substantial proper motion perpendicular to the stream track. The proper motions along the stream ($\Delta \mu_\Lambda \cos B$) are positive, indicating that the stream is moving towards increasing $\Lambda$ in these coordinates. Reassuringly, this is consistent with the observed radial velocity in the Galactic Standard of Rest frame (GSR, $v_{\rm Tuc \, III\, GSR}=-195.2$ km s$^{-1}$, Li et al. in prep.), which shows that the core is moving towards the Galactic center, and the negative radial velocity gradient, which means that the part of the stream with $\Lambda > 0^\circ$ is closer to the Galactic center and hence further along in the direction the stream is moving. 

\subsection{Orbit of Tuc III in the presence of the LMC}

Accounting for the effect of the LMC, we find that Tuc III is on a broadly similar orbit to what we found in \Secref{sec:orbit_mw}. Interestingly, the eccentricity depends on the mass of the LMC, decreasing from an eccentricity of $0.95$, in the presence of a $10^{11}M_\odot$ LMC, to $0.88$ and $0.84$ with a $2\times10^{11}M_\odot$ LMC and a $3\times10^{11}M_\odot$ LMC respectively (see \figref{fig:tuc_orbit}). The least eccentric orbit has a pericenter of $4.8\pm0.8$ kpc, and an apocenter of $54\pm5$ kpc. Furthermore, we find that the orientation of Tuc III's orbital plane also depends on the mass of the LMC. For the more massive LMCs ($M_{\rm LMC} \geq 10^{11}M_\odot$), we find that Tuc III's orbital plane is closely aligned (within $\sim 15^\circ$) with the direction towards the Galactic anti-center, which is close to the orbital pole of the LMC (e.g. see \figref{fig:comparison_figure}). In the presence of less massive LMCs, we find that the orientation of Tuc III's orbit is closely aligned with the direction towards the Galactic center, i.e. almost the opposite direction. The fact that these significantly different orbits can reproduce the same stream (e.g. \figref{fig:comparison_figure}) is equivalent to the fact that the proper motions of Tuc III depend on the mass of the LMC.

\section{Discussion} \label{sec:discussion}

\subsection{Inferring the LMC mass from Tuc III's proper motion} \label{sec:disc_pm}

In this work we have shown that the Tuc III stream can be fit in just the Milky Way potential, as well as in the combined presence of the Milky Way and the LMC. Critically, we find that the expected proper motion of Tuc III depends on the LMC mass. In \Figref{fig:tuc_pm_prediction} we show this proper motion prediction for the LMC masses considered in this work. In principle, a measurement of the proper motion with \textit{Gaia} DR2 would then allow us to infer the mass of the LMC. Li et al. in prep have estimated \textit{Gaia} DR2 can be used to measure the proper motion of Tuc III with a precision of $0.04$ mas yr$^{-1}$. This estimate is based on the 29 stars which Li et al. in prep have confirmed are members of Tuc III based on their radial velocities, metallicities, and location on the color-magnitude diagram. For each star, the \textit{Gaia} G-band magnitude is estimated and then used to compute the expected precision, accounting for the fact that \textit{Gaia} DR2 only makes use of 22 months of data. This precision is shown in \Figref{fig:tuc_pm_prediction,fig:tuc_pm_prediction_LB} and will allow us to measure the effect of the LMC on Tuc III. 

We note that the distance uncertainty to Tuc III is the main contributor to the uncertainty in the proper motions (see \figref{fig:posteriors_MW}). If the distance errors were significantly reduced, this would give a corresponding improvement in the proper motion prediction. \Figref{fig:tuc_pm_prediction_LB} somewhat obscures this since for each realization of the Tuc III stream, we use the actual distance to compute the Sun's reflex proper motion. If we included the uncertainty in the distance, this would give a significantly larger uncertainty in the proper motion perpendicular to the stream, $\mu_B$ (i.e. the large black error bar in \figref{fig:tuc_pm_prediction_LB}). Fortunately, 4 RR Lyrae have recently been found in the Tuc III stream which each have a distance error of $\sim 3\%$ (Mart{\'\i}nez-V{\'a}zquez in prep.). By combining these measurements together, we can get a significantly improved distance to better correct for the Sun's reflex motion and hence measure the misalignment. We assume that the distance error of the combined sample will be 2\%. This improved distance uncertainty would reduce the uncertainty of the Sun's reflex proper motion in $\mu_B$ from roughly 0.17 mas yr$^{-1}$ (10\% distance error) to 0.017 mas yr$^{-1}$ (2\% distance error), which is shown in \Figref{fig:tuc_pm_prediction_LB}. Furthermore, since these RR Lyrae are relatively bright (\textit{Gaia} G $\sim$ 17.5), they will each have a proper motion uncertainty of $\sim0.2$-$0.3$ mas yr$^{-1}$ in \textit{Gaia} DR2, improving the proper motion estimate for Tuc III (Li et al. in prep). We leave the analysis with these RR Lyrae for future work. 

\subsection{Accreting Tuc III with the LMC}

Given the close interaction of Tuc III with the LMC, it is natural to ask whether Tuc III could have accreted along with the LMC. This possibility was included in \Secref{sec:fit_lmc} which found that the Tuc III stream is consistent with a progenitor which has always been disrupting around the Milky Way. However, the technique only allowed for stars to be ejected from Tuc III due to the Milky Way's tidal field. Alternatively, Tuc III could have been tidally disrupting due to the LMC. We explore this possibility by determining the Lagrange points of Tuc III relative to the LMC and stripping particles when Tuc III has a pericenter with respect to the LMC. In this analysis, we also require that Tuc III is bound to the LMC at early times to ensure that Tuc III is physically accreted with the LMC.

We perform a grid search in $\mu_\alpha \cos\delta$ and $\mu_\delta$ as in \Secref{sec:mw_grid}, including a $2\times10^{11}M_\odot$ LMC, but find no fits with likelihoods similar to what was found above. Thus we tentatively conclude that Tuc III is not consistent with having disrupted around the LMC.

\subsection{Limitations of the analysis}

In this work we have assumed a single potential for the Milky Way, namely \texttt{MWPotential2014} from galpy \citep{bovy_galpy}. While this potential satisfies many of the observed constraints on the Milky Way potential \citep{bovy_galpy} and has been shown to be a good fit to some streams around the Milky Way \citep{bovy_pal5_gd1}, our analysis has not accounted for the uncertainties in the Milky Way potential. By testing with other potentials for the Milky Way, namely the best-fit potential in \cite{kuepper_et_al_2015}, we have found that the prediction of the proper motion will depend on the precise potential used. However, these tests revealed the same close passage with the LMC and the same misaligned proper motions. Thus, the more robust prediction is that if the LMC is as massive as expected, it will cause a substantial misalignment between the stream track and proper motions as shown in \Figref{fig:tuc_pm_prediction_LB}. We also stress that in order to use a misaligned proper motion to measure the mass of the LMC, future works will need to marginalize over the uncertainties in the Milky Way potential. 

In this work we have ignored the effect of the Small Magellanic Cloud (SMC). The SMC could potentially influence the orbit of the LMC and exert its own tidal force on the Milky Way. Rotation curve measurements using HI have give a mass of $2.4\times10^9 M_\odot$ within 3 kpc \citep{stanimirovic_smc_mass}. Of course, this is the mass of the SMC within a small aperture; the peak mass (i.e. including dark matter) of the SMC was much larger than this. However, attempts to model the Magellanic stream have found that the SMC has likely had multiple pericentric passages with the LMC \citep[e.g.][]{besla_etal_2012} so much of the SMC's dark matter has already been stripped. Thus, the SMC should have a relatively minor effect on the LMC's orbit. 

Recent works have showed that the Milky Way's bar can affect tidal streams \citep[e.g.][]{hattori_etal_2016,price-whelan_etal_2016,pal5_gaps,pearson_etal_2017}. Since we find that Tuc III ventures very close to the Milky Way center (see \secref{sec:orbit_mw}), we evolve our best-fit Tuc III from \Secref{sec:mw_fit} in the presence of the bar to determine its effect. In order to do this, we take the Milky Way potential described in \Secref{sec:stream_gen_technique}, and replace the bulge with a rotating bar using the analytic model from \cite{long_murali_1992}. For the bar parameters, we take a mass of $5\times10^9 M_\odot$, a length of $a=3$ kpc, a width of $b=1$ kpc, and a present day orientation of $\theta_0=-30^\circ$ \citep[as in][]{hattori_etal_2016,pal5_gaps}. We vary the pattern speed between $0$ and $-100$ km s$^{-1}$ kpc$^{-1}$ in steps of 0.1 km s$^{-1}$ kpc$^{-1}$ (where negative pattern speeds are prograde with the Milky Way disk), more than encompassing the observed constraints on the pattern speed \citep[e.g.][]{bland-hawthorn_gerhard_2016}. We find that the bar has a small effect on the stream, giving slightly different stream tracks depending on the pattern speed of the bar. Thus, the properties of the bar will also need to be marginalized over in order to measure the LMC mass from the misaligned proper motions. 

\subsection{Effect of LMC on other streams in the Milky Way}

The predicted misalignment between the stream track and proper motions should also be seen in any streams which receive a large tidal force from the LMC. Given the orbit of the LMC, we expect that the most affected streams will likely be in the southern Galactic hemisphere, e.g. the 11 streams newly discovered in \cite{shipp_DES_streams}, the Phoenix stream \citep{phoenixdisc}, the Jet stream \citep{jetstreamdisc}, or the ATLAS stream \citep{atlasdisc}. In order to determine which are the most affected, each stream will need to be evolved in the presence of the LMC. 

Interestingly, some works have proposed assuming that stream tracks are aligned with their proper motions to measure the Sun's proper motion \citep{majewski_etal_2006,malhan_ibata_sun_reflex}. If the significant misalignment predicted for the Tuc III stream is verified, this means that these methods must account for the perturbation of the LMC or only use streams which have received a negligible perturbation (possibly those in the northern Galactic hemisphere). 

\subsection{Effect of Tuc III on the Milky Way disk}

Our analysis in \Secref{sec:tuc3_mw_fit} found that Tuc III is on a very eccentric orbit with a pericenter of $\sim1.8$ kpc. This close passage of Tuc III with the Milky Way could induce perturbations in the inner disk of our Galaxy \citep{feldmann_spolyar_2015}. For these perturbations to be significant, Tuc III would have needed a substantial mass of $\sim 10^8-10^9 M_\odot$ and hence would have to be a dwarf galaxy. Given the current data, it is difficult to determine whether Tuc III is a globular cluster or a dwarf (see Li et al. in prep). However, assuming that Tuc III is a dwarf, if Tuc III's recent pericenter was its first approach to the Milky Way (and thus Tuc III still retained the bulk of its dark matter halo) there could be detectable perturbations in the Milky Way center. We note that although we have assumed a disruption age of 3 Gyr, during which our best-fit orbits have experienced multiple pericentric passages (e.g. \figref{fig:tuc_orbit}), our model is not designed to determine when Tuc III began disrupting. Instead, it is designed to match the observed stream track (both on the sky and in radial velocity) and can only be used to set a lower bound on the disruption age needed to make a stream at least as long as Tuc III. Thus, Tuc III having only experienced a single pericentric passage about the Milky Way is not ruled out by our fits.

\section{Conclusions} \label{sec:conclusions}

In this work we have presented the first dynamical stream fits to the Tucana III stream. We stress that this fit assumes a Milky Way potential, namely \texttt{MWPotential2014} from \cite{bovy_galpy}, and gives a prediction of the proper motion of Tuc III. This fit was first done in the presence of only the Milky Way potential which showed that there is a single region of parameter space which gives rise to streams like the Tuc III stream (see \figref{fig:posteriors_MW}). By rewinding the LMC along with Tuc III, the best-fit orbit for Tuc III passes within $\sim 15$ kpc of the LMC within the last 100 Myr. As such, it is crucial to include the LMC when modelling the Tuc III stream.

Including the effect of the LMC on Tuc III, we find that the best-fit streams are indistinguishable given the current data (compare \figref{fig:tuc_MW} and \ref{fig:tuc_2p5e11}). However, the best-fit proper motions are significantly different (\figref{fig:tuc_pm_prediction}). This is because the recent close passage with the LMC exerts a large tidal force on the Tuc III stream which substantially twists the stream. This results in a Tuc III stream whose stream track is significantly misaligned with its proper motion (\figref{fig:tuc_pm_prediction_LB}). Since the proper motion prediction depends on the Milky Way potential, we stress that the misaligned proper motion should be seen as the more robust prediction of this work. 

The upcoming \textit{Gaia} DR2 is expected to revolutionize our understanding of the Milky Way. With astrometric data (sky position, proper motions, and parallax) expected for more than 1.3 billion stars\footnote{\url{https://www.cosmos.esa.int/web/gaia/dr2}}, it should dramatically improve our understanding of the Milky Way's dark matter halo. It will also be of sufficient accuracy to measure the predicted misaligned proper motion of the Tuc III stream. If confirmed, this will be the first direct evidence that the LMC is exerting a substantial perturbation on the Milky Way. Since almost every existing technique for measuring the shape and mass of the Milky Way halo has assumed that the Milky Way is in equilibrium, this perturbation will mean that all of these techniques will need to be revisited. Thus the Tuc III stream may sound the first alarm bells that a precise measurement of the Milky Way halo must account for the LMC. 

\section*{Acknowledgements}

DE thanks Jorge Pe{\~n}arrubia for insightful discussions on the LMC and Victor Debattista for helpful discussions on the bar. DE and VB acknowledge that the research leading to these results has received funding from the European Research Council under the European Union's Seventh Framework Programme (FP/2007- 2013) / ERC Grant Agreement no. 308024. EB acknowledges financial support from the European Research Council (StG-335936). ABP acknowledges generous support from the George P. and Cynthia Woods Institute for Fundamental Physics and Astronomy at Texas A\&M University.

This paper has gone through internal review by the DES
collaboration.

Based on data obtained at the Australian Astronomical Telescope via program A/2016A/26.  We acknowledge the traditional owners of the land on which the AAT stands, the Gamilaraay people, and pay our respects to elders past and present.

Funding for the DES Projects has been provided by the U.S. Department of Energy, the U.S. National Science Foundation, the Ministry of Science and Education of Spain, 
the Science and Technology Facilities Council of the United Kingdom, the Higher Education Funding Council for England, the National Center for Supercomputing 
Applications at the University of Illinois at Urbana-Champaign, the Kavli Institute of Cosmological Physics at the University of Chicago, 
the Center for Cosmology and Astro-Particle Physics at the Ohio State University,
the Mitchell Institute for Fundamental Physics and Astronomy at Texas A\&M University, Financiadora de Estudos e Projetos, 
Funda{\c c}{\~a}o Carlos Chagas Filho de Amparo {\`a} Pesquisa do Estado do Rio de Janeiro, Conselho Nacional de Desenvolvimento Cient{\'i}fico e Tecnol{\'o}gico and 
the Minist{\'e}rio da Ci{\^e}ncia, Tecnologia e Inova{\c c}{\~a}o, the Deutsche Forschungsgemeinschaft and the Collaborating Institutions in the Dark Energy Survey. 

The Collaborating Institutions are Argonne National Laboratory, the University of California at Santa Cruz, the University of Cambridge, Centro de Investigaciones Energ{\'e}ticas, 
Medioambientales y Tecnol{\'o}gicas-Madrid, the University of Chicago, University College London, the DES-Brazil Consortium, the University of Edinburgh, 
the Eidgen{\"o}ssische Technische Hochschule (ETH) Z{\"u}rich, 
Fermi National Accelerator Laboratory, the University of Illinois at Urbana-Champaign, the Institut de Ci{\`e}ncies de l'Espai (IEEC/CSIC), 
the Institut de F{\'i}sica d'Altes Energies, Lawrence Berkeley National Laboratory, the Ludwig-Maximilians Universit{\"a}t M{\"u}nchen and the associated Excellence Cluster Universe, 
the University of Michigan, the National Optical Astronomy Observatory, the University of Nottingham, The Ohio State University, the University of Pennsylvania, the University of Portsmouth, 
SLAC National Accelerator Laboratory, Stanford University, the University of Sussex, Texas A\&M University, and the OzDES Membership Consortium.

Based in part on observations at Cerro Tololo Inter-American Observatory, National Optical Astronomy Observatory, which is operated by the Association of 
Universities for Research in Astronomy (AURA) under a cooperative agreement with the National Science Foundation.

The DES data management system is supported by the National Science Foundation under Grant Numbers AST-1138766 and AST-1536171.
The DES participants from Spanish institutions are partially supported by MINECO under grants AYA2015-71825, ESP2015-66861, FPA2015-68048, SEV-2016-0588, SEV-2016-0597, and MDM-2015-0509, 
some of which include ERDF funds from the European Union. IFAE is partially funded by the CERCA program of the Generalitat de Catalunya.
Research leading to these results has received funding from the European Research
Council under the European Union's Seventh Framework Program (FP7/2007-2013) including ERC grant agreements 240672, 291329, and 306478.
We  acknowledge support from the Australian Research Council Centre of Excellence for All-sky Astrophysics (CAASTRO), through project number CE110001020, and the Brazilian Instituto Nacional de Ci\^encia
e Tecnologia (INCT) e-Universe (CNPq grant 465376/2014-2).

This manuscript has been authored by Fermi Research Alliance, LLC under Contract No. DE-AC02-07CH11359 with the U.S. Department of Energy, Office of Science, Office of High Energy Physics. The United States Government retains and the publisher, by accepting the article for publication, acknowledges that the United States Government retains a non-exclusive, paid-up, irrevocable, world-wide license to publish or reproduce the published form of this manuscript, or allow others to do so, for United States Government purposes.

\bibliographystyle{mn2e_long}
\bibliography{citations_tuc3}

\section*{Affiliations}

$^{1}$ Department of Physics, University of Surrey, Guildford GU2 7XH, UK\\
$^{2}$ Institute of Astronomy, University of Cambridge, Madingley Road, Cambridge CB3 0HA, UK\\
$^{3}$ Fermi National Accelerator Laboratory, P. O. Box 500, Batavia, IL 60510, USA\\
$^{4}$ Kavli Institute for Cosmological Physics, University of Chicago, Chicago, IL 60637, USA\\
$^{5}$ McWilliams Center for Cosmology, Department of Physics, Carnegie Mellon University, 5000 Forbes Avenue, Pittsburgh, PA 15213, USA\\
$^{6}$ Center for Computational Astrophysics, Flatiron Institute, 162 5th Avenue, New York, NY 10010, USA\\
$^{7}$ LSST, 933 North Cherry Avenue, Tucson, AZ 85721, USA\\
$^{8}$ Department of Physics, University of Illinois at Urbana-Champaign, 1110 W. Green St., Urbana, IL 61801, USA\\
$^{9}$ Australian Astronomical Observatory, North Ryde, NSW 2113, Australia\\
$^{10}$ George P. and Cynthia Woods Mitchell Institute for Fundamental Physics and Astronomy, and Department of Physics and Astronomy, Texas A\&M University, College Station, TX 77843,  USA\\
$^{11}$ Cerro Tololo Inter-American Observatory, National Optical Astronomy Observatory, Casilla 603, La Serena, Chile\\
$^{12}$ Observatories of the Carnegie Institution for Science, 813 Santa Barbara St., Pasadena, CA 91101, USA\\
$^{13}$ Department of Physics, Stanford University, 382 Via Pueblo Mall, Stanford, CA 94305, USA\\
$^{14}$ Kavli Institute for Particle Astrophysics \& Cosmology, P. O. Box 2450, Stanford University, Stanford, CA 94305, USA\\
$^{15}$ SLAC National Accelerator Laboratory, Menlo Park, CA 94025, USA\\
$^{16}$ Department of Physics \& Astronomy, University College London, Gower Street, London, WC1E 6BT, UK\\
$^{17}$ Department of Physics and Electronics, Rhodes University, PO Box 94, Grahamstown, 6140, South Africa\\
$^{18}$ Institute of Cosmology \& Gravitation, University of Portsmouth, Portsmouth, PO1 3FX, UK\\
$^{19}$ CNRS, UMR 7095, Institut d'Astrophysique de Paris, F-75014, Paris, France\\
$^{20}$ Sorbonne Universit\'es, UPMC Univ Paris 06, UMR 7095, Institut d'Astrophysique de Paris, F-75014, Paris, France\\
$^{21}$ Laborat\'orio Interinstitucional de e-Astronomia - LIneA, Rua Gal. Jos\'e Cristino 77, Rio de Janeiro, RJ - 20921-400, Brazil\\
$^{22}$ Observat\'orio Nacional, Rua Gal. Jos\'e Cristino 77, Rio de Janeiro, RJ - 20921-400, Brazil\\
$^{23}$ Department of Astronomy, University of Illinois at Urbana-Champaign, 1002 W. Green Street, Urbana, IL 61801, USA\\
$^{24}$ National Center for Supercomputing Applications, 1205 West Clark St., Urbana, IL 61801, USA\\
$^{25}$ Institut de F\'{\i}sica d'Altes Energies (IFAE), The Barcelona Institute of Science and Technology, Campus UAB, 08193 Bellaterra (Barcelona) Spain\\
$^{26}$ Department of Physics and Astronomy, University of Pennsylvania, Philadelphia, PA 19104, USA\\
$^{27}$ Centro de Investigaciones Energ\'eticas, Medioambientales y Tecnol\'ogicas (CIEMAT), Madrid, Spain\\
$^{28}$ Department of Astronomy/Steward Observatory, 933 North Cherry Avenue, Tucson, AZ 85721-0065, USA\\
$^{29}$ Jet Propulsion Laboratory, California Institute of Technology, 4800 Oak Grove Dr., Pasadena, CA 91109, USA\\
$^{30}$ Department of Astronomy, University of Michigan, Ann Arbor, MI 48109, USA\\
$^{31}$ Department of Physics, University of Michigan, Ann Arbor, MI 48109, USA\\
$^{32}$ Instituto de Fisica Teorica UAM/CSIC, Universidad Autonoma de Madrid, 28049 Madrid, Spain\\
$^{33}$ Institut d'Estudis Espacials de Catalunya (IEEC), 08193 Barcelona, Spain\\
$^{34}$ Institute of Space Sciences (ICE, CSIC),  Campus UAB, Carrer de Can Magrans, s/n,  08193 Barcelona, Spain\\
$^{35}$ Department of Physics, ETH Zurich, Wolfgang-Pauli-Strasse 16, CH-8093 Zurich, Switzerland\\
$^{36}$ Santa Cruz Institute for Particle Physics, Santa Cruz, CA 95064, USA\\
$^{37}$ Center for Cosmology and Astro-Particle Physics, The Ohio State University, Columbus, OH 43210, USA\\
$^{38}$ Department of Physics, The Ohio State University, Columbus, OH 43210, USA\\
$^{39}$ Harvard-Smithsonian Center for Astrophysics, Cambridge, MA 02138, USA\\
$^{40}$ Instituci\'o Catalana de Recerca i Estudis Avan\c{c}ats, E-08010 Barcelona, Spain\\
$^{41}$ Instituto de F\'\i sica, UFRGS, Caixa Postal 15051, Porto Alegre, RS - 91501-970, Brazil\\
$^{42}$ School of Physics and Astronomy, University of Southampton,  Southampton, SO17 1BJ, UK\\
$^{43}$ Brandeis University, Physics Department, 415 South Street, Waltham MA 02453\\
$^{44}$ Instituto de F\'isica Gleb Wataghin, Universidade Estadual de Campinas, 13083-859, Campinas, SP, Brazil\\
$^{45}$ Computer Science and Mathematics Division, Oak Ridge National Laboratory, Oak Ridge, TN 37831\\

\end{document}